\documentclass[twocolumn,showpacs,showkeys,amsmath,amssymb,prb]{revtex4}
\usepackage{graphicx}

\begin{document}

\title{Anisotropy of spin-orbit induced electron spin relaxation in [001] and [111] grown GaAs quantum dots}

\author{C. Segarra}
\author{J. Planelles}
\email{josep.planelles@uji.es}
\author{J. I. Climente}
\author{F. Rajadell}
\affiliation{Departament de Qu\'{\i}mica F\'{\i}sica i Anal\'{\i}tica, Universitat Jaume I, Castell\'o de la Plana, Spain}

\date{\today}


\begin{abstract}

We report a systematic study of the spin relaxation anisotropy between single electron Zeeman sublevels in 
cuboidal GaAs quantum dots (QDs).    The QDs are subject to an in-plane magnetic field. 
As the field orientation varies, the relaxation rate oscillates periodically, showing ``magic'' angles
where the relaxation rate is suppressed by several orders of magnitude.
This behavior is found in QDs with different shapes, heights, crystallographic orientations and external fields.
The origin of these angles can be traced back to the symmetries of the spin admixing terms of the Hamiltonian.
In [001] grown QDs, the suppression angles are different for Rashba and Dresselhaus spin-orbit terms. 
By contrast, in [111] grown QDs they are the same, which should facilitate a thorough suppression of 
spin-orbit induced relaxation.
Our results evidence that cubic Dresselhaus terms play a critical role in determining the
spin relaxation anisotropy even in quasi-2D QDs.


\end{abstract}


 
\maketitle

\section{Introduction}

The electron spin confined in semiconductor QDs is a promising candidate for the realization of quantum computing and the development of spin-based devices in spintronics.\cite{Awschalom2002,Fabian2007}
Using the spin of electrons as qubits was first proposed by Loss and DiVincenzo (Ref. \onlinecite{Loss1998}) and, since then, a lot of effort has been devoted to its accomplishment.\cite{Hanson2007} 
QDs offer the possibility of isolating single electron spins which exhibit longer lifetimes than in delocalized systems since quantum confinement suppresses the main bulk decoherence mechanisms.\cite{khaetskii2000} 
Nevertheless, coupling between the electron spin and the surrounding environment cannot be avoided, resulting in spin relaxation and decoherence. 
Therefore, a good understanding of the relaxation mechanisms in QDs is needed for the development of spin-based applications.

The two main mechanisms of spin relaxation in III-V zinc-blende semiconductor QDs are the hyperfine coupling with the nuclear spins of the lattice and the spin-orbit interaction (SOI).\cite{Hanson2007}
The hyperfine interaction is generally important at relatively weak magnetic fields while for moderate and strong fields the phonon-mediated relaxation due to SOI predominates.
In semiconductors without inversion symmetry, e.g. GaAs, SOI can be originated by the bulk inversion asymmetry of the material (Dresselhaus SOI)\cite{Dress1955} and the structure inversion asymmetry of the confining potential (Rashba SOI).\cite{Rashba1984}
The Hamiltonians describing both SOI have different symmetries and exhibit an anisotropic behavior.\cite{Winkler2003}
This anisotropy can be exploited to externally control and manipulate the electron spin by changing the orientation 
of applied magnetic or electric fields.\cite{Konemann2005,Takahashi2010,Nowak2013}
As a consequence, the anisotropy of the spin relaxation and its control via external means has been intensively studied.\cite{Falko2005,Destefani2005,Stano2006,Olendski2007,Amasha2008,Prabhakar2013} 

Most works so far have dealt with two-dimensional (2D) parabolic InAs or GaAs QDs grown along the [001] crystal 
direction,\cite{Hanson2007,Falko2005,Destefani2005,Stano2006} where in-plane anisotropy arises from the interference between 
Rashba and Dresselhaus SOI.
However, realistic QDs are prone to deviate from the circular symmetry and there is gathering evidence that this has 
a primary influence on the spin relaxation anisotropy.\cite{Olendski2007,Amasha2008,Prabhakar2013} 
Yet, previous works on this aspect neglected cubic Dresselhaus SOI terms, whose role may be important.
Cubic terms are expected to become particularly important in QDs with large height-to-base aspect ratio\cite{Planelles2012},
which are increasingly available owing to recent progress in synthetic control.\cite{Dalacu2011,Pfund2006}
Going beyond [001] grown QDs is also of interest, especially in view of the convenience of [111] grown QDs 
for optical spin preparation.\cite{Mano2010}
 The effect of the crystallographic orientation on the spin dynamics has been well studied in quantum 
wells\cite{Vurgaftman2005,Biermann2012,Balocchi2013}, but further work is needed in relation to fully localized spins.

In this work, we study the anisotropy of the electron spin relaxation between Zeeman sublevels in cuboidal GaAs QDs.
The anisotropy is monitored by varying the orientation of an externally applied in-plane magnetic field ($\phi_B$).
We consider QDs grown along both [001] and [111] crystal directions, including all linear and cubic terms
of Rashba and Dresselhaus SOI in a fully 3D model. Different heights, base shapes, crystallographic orientations
and external electric fields are considered. The numerical results, together with perturbative interpretations, 
provide a wide overview on the effect of confinement asymmetry and three-dimensionality 
on the spin relaxation anisotropy.

We find that, in [001] grown QDs, the spin relaxation anisotropy is very different depending on the dominating
spin-orbit mechanism, Rashba or Dresselhaus SOI. By contrast, in [111] grown QDs the anisotropy is the same for
both terms. In all cases, the spin relaxation rate shows strong oscillations with $\phi_B$. 
Interestingly, cubic Dresselhaus terms are shown to be critical in determining such anisotropic behavior. 
This occurs not only in high QDs, but -- contrary to common belief -- also in quasi-2D QDs, provided the high 
symmetry directions of the dot are not aligned with the main crystallographic axes.  
In both squared and rectangular QDs we observe order-of-magnitude suppressions of the spin relaxation rate
at certain ``magic'' magnetic field angles $\phi_B$, which can be understood from symmetry considerations.

The paper is organized as follows. 
Sec. II presents the model we use to compute the electron spin relaxation, including the SOI Hamiltonians for
QDs rotated with respect to the main crystallographic axes.
In Sec. III we show and discuss the numerical results for the cases under study.
Finally, conclusions are given in Sec. IV.

\section{Theoretical model}

\begin{figure}[h!]
\includegraphics[width=0.35\textwidth]{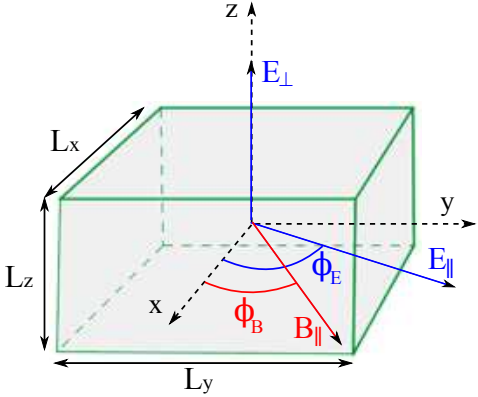}
\caption{Schematic representation of the cuboidal QD system. The orientation of the external electric and magnetic fields is indicated.}
\label{fig1}
\end{figure}
 
We study the electron spin relaxation driven by SOI between Zeeman split sublevels of cuboidal GaAs QDs subject to externally applied electric $\mathbf{E}$ and magnetic $\mathbf{B}$ fields (see Fig.~\ref{fig1}). 
The one-electron states are described by a three-dimensional Hamiltonian of the form
\begin{equation}
\label{eq1}
H =  \frac{\mathbf{p}^{2}}{2m^{*}}+V_{c}+\mathbf{E}\,\mathbf{r} +H_{Z}+H_{SOI},
\end{equation}
\noindent where $m^{*}$ stands for the electron effective mass, $V_{c}$ is the confinement potential, $\mathbf{E}$ is an external electric field and $\mathbf{p}=-i\hbar\boldsymbol{\nabla}+\mathbf{A}$,
where $\mathbf{A}$ is the vector potential.
An in-plane magnetic field $\mathbf{B}=B\left(\cos\phi_B,\sin\phi_B,0\right)$ rotated an angle $\phi_B$ with respect to the $x$ axis of the dot is included.
This field is described by the  vector potential $\mathbf{A}=\left( z B \sin \phi_B, -z B \cos\phi_B,0 \right)$.
The Zeeman term is $H_{Z}=\frac{1}{2}g\mu_{B}\mathbf{B}\,\boldsymbol{\sigma}$ with $g$, $\mu_B$ and $\boldsymbol{\sigma}$ standing for the electron g-factor, Bohr magneton and Pauli spin matrices, respectively.

The last term in Eq. \eqref{eq1} corresponds to the SOI,\cite{Winkler2003} $H_{SOI}=H_{R}+H_{D}$, with $H_{R}$ being the Rashba SOI
\begin{equation}
\label{eqrashba001}
H_{R}^{[001]}=r\boldsymbol{\sigma}\left( \mathbf{p} \times \mathbf{E} \right),
\end{equation}
\noindent and $H_{D}$ the Dresselhaus SOI
\begin{equation}
\label{eqdress001}
\resizebox{0.9 \hsize}{!}{$H_{D}^{[001]}=d \left[\sigma_{x} p_{x}\left( p_{y}^{2} - p_{z}^{2} \right)+\sigma_{y} p_{y}\left( p_{z}^{2} - p_{x}^{2} \right)+\sigma_{z} p_{z}\left( p_{x}^{2} - p_{y}^{2} \right) \right]$}
\end{equation}
\noindent Here, $r$ and $d$ are material-dependent coefficients determining the strength of the SOI and the superscript $[001]$ indicates de growth direction of the QD.

Eq. \eqref{eqrashba001} and Eq. \eqref{eqdress001} correspond to QDs grown along the [001] crystal direction.
In order to consider other orientations of the QD with respect to the crystal host we maintain the confinement potential fixed in space and perform a rotation of the crystalline structure.
Since the confining potential as well as the externally applied fields are kept while the crystalline structure is rotated, only the $H_{SOI}$ part of the Hamiltonian is affected. In particular,
the $H_{SOI}$  Hamiltonian corresponding to an axially applied electric field and a crystalline structure subject to an in-plane rotation $\theta_z$ around the $z$ axis read:
\begin{equation}
\label{rashrotz}
H_{R}^{[001]} (\theta_z)= r E_{z}  (\sigma_x p_y - \sigma_y p_x ),
\end{equation}
\noindent and
\begin{equation}
\label{dressrotz}
\begin{split}
H_{D}^{[001]} &(\theta_z)= d \cos 2\theta_z \big[ \sigma_{x} p_{x}\left( p_{y}^{2} - p_{z}^{2} \right) +\sigma_{y} p_{y}\left( p_{z}^{2} - p_{x}^{2} \right)\\
&+\sigma_{z} p_{z}\left( p_{x}^{2} - p_{y}^{2} \right) \big] +d \sin 2\theta_z \big[ p_z^2 (\sigma_y p_x+ \sigma_x p_y)\\
&-2\sigma_z p_x p_y p_z+ \frac{1}{2} (p_x^2 - p_y^2) (\sigma_x p_y - \sigma_y p_x) \big].
\end{split}
\end{equation}
\noindent Note that this particular case of an axially applied electric field yields a Rashba Hamiltonian \eqref{rashrotz} independent of $\theta_z$.\\

We consider next QDs grown along the [111] direction. In particular, we consider the rotation $\chi=\arccos (1/\sqrt{3})$ around the straight line $y=-x$, that correspond to the Euler angles $\theta= \arccos (1/\sqrt{3})$, $\phi=45$ and $\alpha=-45$. The rotated SOI Hamiltonians have the form
\begin{equation}
\label{rash111}
H^{[111]}_{R}= \frac{r\:E_z}{\sqrt 3} \left[ \sigma_z (p_y-p_x)- \sigma_y (p_x+p_z) + \sigma_x (p_y+p_z) \right],
\end{equation}
\noindent and
\begin{equation}
\label{dress111}
\begin{split}
	H^{[111]}_{D}&= \frac{d}{2\sqrt 3} [ (p_x^2 + p_y^2 -4 p_z^2)(p_x \sigma_y - p_y \sigma_x) \\
	&+ p_z (p_x^2 -p_y^2)(\sigma_x+\sigma_y)+2 p_x p_y p_z (\sigma_x-\sigma_y) \\
        &-\sigma_z p_x^2 (p_x+3 p_y)+\sigma_z p_y^2 (p_y+3 p_x) ],
\end{split}
\end{equation}
\noindent where the electric field is aligned with the dot $z$ axis.

The relaxation rate between the initial electron state $|\Psi_{i}\rangle$ and the final electron state $|\Psi_{f}\rangle$ 
is estimated by the Fermi golden rule
\begin{equation}
\label{fermieq}
\frac{1}{T_{1}} = \frac{2\pi}{\hbar} \sum_{\lambda,\mathbf{q}} \left| M_{\lambda} (\mathbf{q}) \right|^{2} 
\left| \langle\Psi_{f} \left| e^{-i \mathbf{q} \mathbf{r}} \right| \Psi_{i} \rangle\right|^2 \delta (E_f - E_i - E_q )
\end{equation}
Here, $\mathbf{q}$ is the bulk phonon wave vector and $M_{\lambda}(\mathbf{q})$ denotes the scattering matrix element corresponding to the electron-phonon interaction $\lambda$, being the piezoelectric or the deformation potentials.\cite{Climente2006}
All calculations are carried out at zero temperature, thus only phonon emission processes are possible. 
The splitting energy between Zeeman sublevels is small so that only acoustic phonons are important and the linear dispersion regime applies $E_{q} = \hbar c_{\alpha} q$, where $c_{\alpha}$ is the sound velocity of the longitudinal or transversal phonon branch.\cite{khaetskii2001}
Note that phonons cannot couple states with opposite spin and the spin admixture caused by SOI is essential for relaxation to take place.

The eigenvalue problem is solved numerically using a finite difference method on a three-dimensional grid.
Accounting for SOI in the calculation of the energy spectra requires high numerical precision due to the small magnitude of this coupling and the presence of third-order derivatives.
Different approaches to approximate the derivatives in the finite difference method have been studied.
After a series of convergence tests, a 7-point stencil central difference scheme and a number of 42875 mesh nodes discretizing the 3D system has been employed in all calculations yielding matrices of dimensions 85750x85750.
With this, we guarantee an accurate description of the electron states at a reasonable computational cost.

The QD system is described with a hard-wall confinement potential. 
We use GaAs material parameters, particularly electron effective mass $m^{*}=0.067$, density $\rho=5310\:kg/m^3$, dielectric constant $\epsilon_{r}=12.9$, piezoelectric constant $h_{14}=1.45\cdot10^9\:V/m$, g-factor $g=-0.44$ and sound velocities $c_l=4720\:m/s$ and $c_t=3340\:m/s$.\cite{Vurgaftman2005,LB-Madelung}
For the SOI constants, we take $d=27.58\:eV\AA^3$ and $r=5.026\:e\AA^2$.\cite{Winkler2003}
All simulations are carried out, unless otherwise stated, considering an axial electric field $E_z=10\:kV/cm$ and an in-plane magnetic field $B_{\parallel}=1T$.

\section{Results and discussion}
\subsection{Geometry dependence}

We investigate first the relaxation rate anisotropy for different dot geometries when applying an in-plane magnetic field at different orientations.
The QDs considered have a base with square ($L_x=80\:nm$, $L_y=80\:nm$) or rectangular ($L_x=70\:nm$, $L_y=90\:nm$) shape and various heights ranging from $L_z=10\:nm$ to $L_z=40\:nm$.

\begin{figure}[h]
\includegraphics[width=0.45\textwidth]{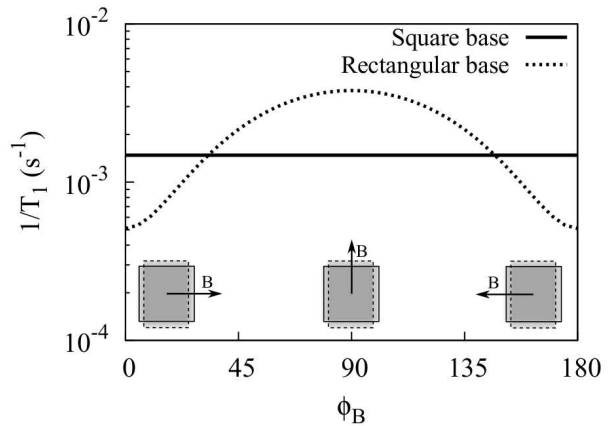}
\caption{Electron spin relaxation rate as a function of the in-plane magnetic field orientation when only the Rashba SOI contribution is included. QDs of 10nm height with rectangular (dotted line) and square base (solid line) are considered.}
\label{fig2}
\end{figure}

Fig.~\ref{fig2} shows the spin relaxation rate when only Rashba SOI is present.
For QDs with square base the relaxation rate is the same regardless of $\phi_B$.
In contrast, in rectangular QDs it presents an anisotropic behavior, where the maximum (minimum) corresponds to a magnetic field oriented along the direction of weaker (stronger) confinement.
In both cases, $1/T_{1}$ is independent of the QD height and, for the sake of clarity, only results for $L_z=10\:nm$ are included in Fig.~\ref{fig2}. 

\begin{figure}[h]
\includegraphics[width=0.45\textwidth]{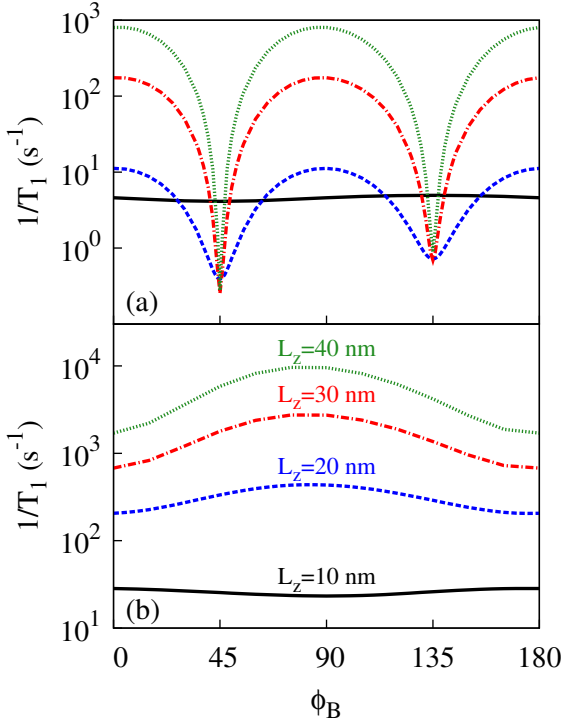}
\caption{Calculated spin relaxation rate vs. magnetic field orientation $\phi_B$ considering only Dresselhaus SOI in (a) square and (b) rectangular base QDs. Different dot heights are studied: $L_z=10\:nm$ (solid black line), $L_z=20\:nm$ (blue dashed line), $L_z=30\:nm$ (red dash-dotted line) and $L_z=40\:nm$ (green dotted line).}
\label{fig3}
\end{figure}

In Fig.~\ref{fig3}(a), we analyze the spin relaxation in the only presence of Dresselhaus SOI for QDs with square base.
The relaxation rate for small QDs ($L_z=10\:nm$) is almost isotropic with the orientation of the magnetic field.
This is in sharp contrast with higher QDs, where strong quenchings are found at $\phi_B=45$ and $\phi_B=135$. 
On the other hand, when the QD base is rectangular, Fig.~\ref{fig3}(b), only moderate modulations of $1/T_{1}$ are observed.
Again, the dependence on $\phi_B$ is different depending on the dot height. 
When $B_{\parallel}$ is oriented along the direction of weaker confinement the relaxation is minimum for QDs with $L_z=10\:nm$,
but it changes into a maximum for $L_z=20,30,40\:nm$.

The preceding results reveal a strong sensitivity of the spin relaxation anisotropy to both the QD symmetry 
(squared or rectangular) and the QD height. Both factors can induce major, qualitative changes in the anisotropy.
To understand such a behavior, we consider that the relaxation rate is proportional to the degree of spin admixture
of the initial and final states of the transition, $\Psi_i$ and $\Psi_f$ in Eq.~(\ref{fermieq}).\cite{khaetskii2001}
These states can be approximated as: 
\begin{equation}
\begin{split}
\Psi_i \approx \psi_{000} |\!\downarrow \rangle + c_x^i \psi_{100} |\!\uparrow \rangle + c_y^i \psi_{010} |\!\uparrow \rangle \\
\Psi_f \approx \psi_{000} |\!\uparrow \rangle + c_x^f \psi_{100}|\!\downarrow \rangle + c_y^f \psi_{010} |\!\downarrow \rangle
\label{eqstates}
\end{split}
\end{equation}
\noindent where $\psi_{ijk}$ represents the electron orbital in the absence of SOI, with ${ijk}$ the number of nodes in $x$, $y$ and $z$,
respectively, while $|\!\uparrow \rangle$ ($|\!\downarrow \rangle$) represents parallel (antiparallel) spin alignment along the direction 
of the magnetic field.
For the analysis we can focus on $\Psi_i$ (analogous reasoning is valid for $\Psi_f$).
$\Psi_i$ is mostly a spin down state, with a little SOI induced spin admixture with excited levels.
 Notice that $\psi_{000} |\!\uparrow\rangle $ does not contribute to the spin admixture of $\Psi_i$ 
because the parity symmetry in $x$ and $y$ prevents direct SOI coupling with $\psi_{000} |\!\downarrow\rangle$ .
Thus, the degree of spin admixture is essentially captured by the coefficients $c_x^i$ and $c_y^i$, 
which can be estimated perturbatively as:
\begin{equation}
c_x^i = - \frac{ \langle \uparrow |  \langle \psi_{100} | H_{SOI} | \psi_{000} \rangle \, |\!\downarrow \rangle }{\varepsilon_{100\uparrow}-\varepsilon_{000\downarrow}},
\end{equation}
\noindent and 
\begin{equation}
c_y^i = - \frac{ \langle \uparrow |  \langle \psi_{010} | H_{SOI} | \psi_{000} \rangle \, |\!\downarrow \rangle }{\varepsilon_{010\uparrow}-\varepsilon_{000\downarrow}}.
\end{equation}
\noindent 
The energy separations $\Delta \varepsilon_x = \varepsilon_{100\uparrow}-\varepsilon_{000\downarrow}$ and $\Delta \varepsilon_y = \varepsilon_{010\uparrow}-\varepsilon_{000\downarrow}$
do not vary with $\phi_B$. Thus, the origin of the anisotropy must be sought in the SOI matrix elements. 

We consider first Rashba SOI, i.e. $H_{SOI}=H_R^{[001]}(0)$. From Eq.~(\ref{rashrotz}) and parity considerations,
it follows that, for $\phi_B=0$,
\begin{align}
c_x^i &= r\,E_z \frac{\langle \uparrow | \sigma_y | \downarrow \rangle \langle \psi_{100} | p_x | \psi_{000}\rangle}{\Delta \varepsilon_x}, &
c_y^i &= 0
\end{align}
\noindent while for $\phi_B=90$,
\begin{align}
c_x^i & = 0, &
c_y^i &= r\,E_z \frac{\langle \uparrow | \sigma_x | \downarrow \rangle \langle \psi_{010} | p_y | \psi_{000}\rangle}{\Delta \varepsilon_y}.
\end{align}
We see that depending on the orientation of the magnetic field the spin admixture is caused by the coupling to a different excited state.
For QDs with square base $\Delta \varepsilon_x = \Delta \varepsilon_y$, and $\langle \psi_{100} | p_x | \psi_{000} \rangle = \langle \psi_{010} | p_y | \psi_{000}\rangle$.
Consequently, the degree of spin mixing does not change at $\phi_B=0$ and $\phi_B=90$, in agreement with the isotropic $1/T_1$ observed in Fig.~\ref{fig2}.
Conversely, in rectangular QDs with stronger confinement in $x$, $\Delta \varepsilon_x > \Delta \varepsilon_y$. 
Then, the admixture coefficients at $\phi_B=90$ are larger than at $\phi_B=0$, which justifies the anisotropy observed in Fig.~\ref{fig2}.

The anisotropy of Dresselhaus SOI induced spin relaxation, shown in Fig.~\ref{fig3}, can be understood in similar terms.
We split Eq.~\eqref{eqdress001} as $H_{D}^{[001]}=H_z+H_{xy}$, where $H_{z}=d\:p_z^2 \left( p_y \sigma_y - p_x \sigma_x \right)$ 
and $H_{xy}=H_{x}+H_{y}=d \left[ p_x^2 \left( p_z \sigma_z - p_y \sigma_y \right)+p_y^2 \left( p_x \sigma_x - p_z \sigma_z \right) \right]$.
Calculations using these Hamiltonians independently show that $H_{z}$ dominates for $L_z=10\:nm$, in agreement with the usual practice of approximating the Dresselhaus SOI by $H_{z}$ in quasi-2D systems. 
If we perform a similar analysis for $H_z$ as the one carried out for Rashba SOI, we find that coupling to $\psi_{010}$ and $\psi_{100}$ dominates at $\phi_B=0$ and $\phi_B=90$, respectively.
This is exactly the opposite as for the Rashba SOI case, explaining the results obtained for $L_z=10\:nm$ QDs (see Fig.~\ref{fig3}(b)).
As the QD height is increased, however, $H_{xy}$ soon dominates over $H_z$. Indeed, for $L_z=20\:nm$ it is already dominant. 
Considering individually $H_x$ and $H_y$ it can be shown that they present opposite behaviors with $\phi_B$.
$H_x$ produces a maximum (minimum) relaxation for $\phi_B=90$ ($\phi_B=0$) and $H_x$ for $\phi_B=0$ ($\phi_B=90$).
This dependence does not change with the base shape and a stronger confinement in one direction only determines which term, 
$H_x$ or $H_y$, prevails. In the rectangular dot of Fig.~\ref{fig3}(b), $L_x<L_y$ so $H_x$ is more important and we 
observe its angular dependence.  
Instead, when the dot base is squared $H_x$ and $H_y$ cancel each other out at $\phi_B=45$ and $\phi_B=135$, thus 
giving rise to the pronounced minima of $1/T_{1}$ observed in Fig.~\ref{fig3}(a).

To summarize this section, the spin relaxation anisotropy of [001] grown GaAs QDs is determined by the spin admixture induced by SOI.
This is qualitatively different in systems where Rashba or Dresselhaus SOI terms dominate. In the latter case,
the anisotropy reflects whether $H_z$ or $H_{xy}$ prevails. It turns out that $H_{xy}$ is already dominant for $L_z=20\:nm$ 
(height-to-base aspect ratio of 1:4), which points out at the early relevance of cubic Dresselhaus terms 
in structures where three-dimensionality starts becoming important. In this case, the use of QDs with symmetric x-y
confinement enables strong suppressions of the relaxation at certain magnetic field orientations.
These ``magic'' angles are reminiscent of the ``easy passages'' found by Stano and Fabian for laterally
coupled circular QDs.\cite{Stano2006} In this and the following sections we show that related
physics arises for single QDs with non-circular confinement, which has significant practical implications.

\subsection{In-plane confinement potential orientation}

\begin{figure}[h!]
\includegraphics[width=0.45\textwidth]{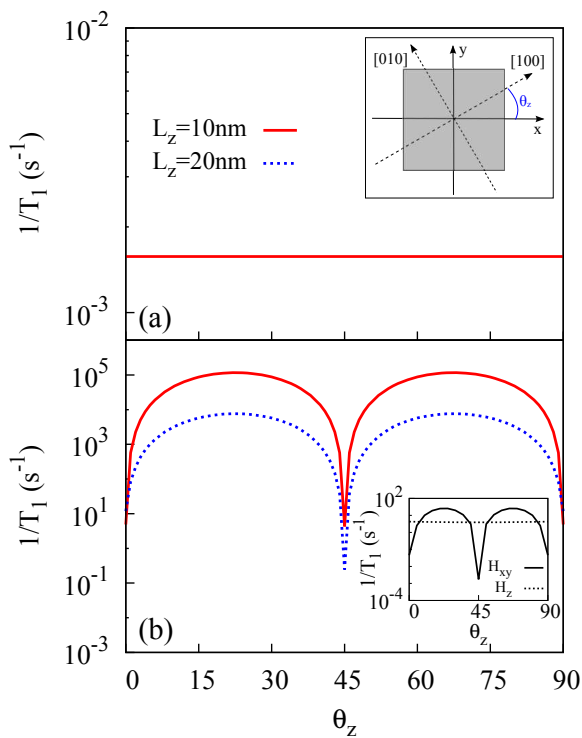}
\caption{Spin relaxation rate as a function of the dot orientation $\theta_z$ for square base QDs with $L_z=10\:nm$ (black solid curve) and $L_z=20\:nm$ (blue dotted curve). Results are shown for (a) pure Rashba SOI and (b) pure Dresselhaus SOI. The in-plane magnetic field $B_{\parallel}=1\:T$ is oriented along the dot $x$ axis ($\phi_B=0$). The inset in (a) illustrates a representation of the system and the definition of the rotation angle. The inset in (b) shows the relaxation due to $H_{xy}$ and $H_z$ in the $L_z=10\:nm$ dot.}
\label{fig4}
\end{figure}

In this section, we investigate the impact of the QD orientation with respect to the crystal host on the spin relaxation.
 The rotation angle $\theta_z$ is defined as the angle between the [001] crystal direction and the $x$ axis of the dot, see inset of Fig.~\ref{fig4} for a schematic representation.
All calculations are carried out with the magnetic field $B_{\parallel}=1\:T$ oriented along the $x$ axis of the QD and an axial electric field $E_z=10\:kV/cm$.

In Fig.~\ref{fig4}(a), we plot the relaxation rate in the presence of Rashba SOI only for QDs with $L_z=10\:nm$ (results for $L_z=20\:nm$ are identical and are omitted for clarity).
We find that $1/T_{1}$ is not affected by changes in the dot orientation.
This result is as expected since Eq.~\eqref{rashrotz} does not depend on $\theta_z$.

For the Dresselhaus SOI case instead, Fig.~\ref{fig4}(b) shows a strong dependence of $1/T_{1}$ on the confinement potential rotation.
In particular, one can see some specific rotation angles, $\theta_z=0,45,90$, where the spin relaxation is reduced by 4-5 orders of magnitude
as compared to others. This behavior can be understood from the form of the Hamiltonian in Eq.~\eqref{dressrotz}.
The Dresselhaus SOI presents a $2\theta_z$ dependence, with half of the terms multiplied by $\sin 2\theta_z$ 
and the other half by $\cos 2\theta_z$. Therefore, the first part of Eq.~\eqref{dressrotz} cancels for $\theta_z=45$ 
and the second part for $\theta_z=0$ and $\theta_z=90$. This suppresses some of the SOI coupling channels,
giving rise to slower relaxation rate than for intermediate angles.

It is noteworthy to mention that the dependence on $\theta_z$ originates in $H_{xy}$, with $H_z$ remaining isotropic, 
 see Fig.~\ref{fig4}(b) inset. This highlights the important role of the cubic terms of the Dresselhaus SOI Hamiltonian in GaAs QDs. 
As a matter of fact, the inset shows that even in the shortest QDs ($L_z=10\:nm$), save for the vicinity of the ``magic'' 
rotation angles ($\theta_z=0,\,45,\,90$) the main contribution to the relaxation rate does not come from $H_z$ but from $H_{xy}$.

These results are robust against changes in the QD geometry, such as height and base shape, which do not modify the qualitative trend.
In particular, the minimum at $\theta_z=45$ remains unaltered while the minima at $\theta_z=0$ and $\theta_z=90$ is slightly shifted in rectangular QDs. 

\subsection{Effect of an additional in-plane electric field}

We next explore the influence of applying an in-plane electric field on the spin relaxation anisotropy.
We consider the squared QD of Sec. IIIA with $B_{\parallel}=1\:T$ and $E_z=10\:kV/cm$, but now we add 
an additional electric field component $E_{\parallel}=10\:kV/cm$.
Calculations are performed rotating the in-plane electric field for some fixed magnetic field orientations.

\begin{figure}[h!]
\includegraphics[width=0.45\textwidth]{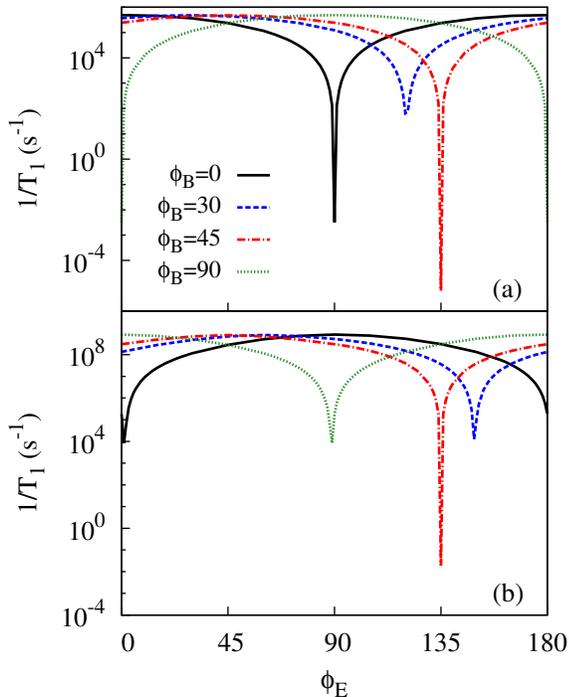}
\caption{Electron spin relaxation as a function of the in-plane electric field orientation $\phi_E$ considering (a) only Rashba SOI and (b) only Dresselhaus SOI. The QDs studied have square base and $L_z=10\:nm$. Calculations with the magnetic field oriented at some fixed angles are presented: $\phi_B=0$ (black solid line), $\phi_B=30$ (blue dashed line), $\phi_B=45$ (red dash-dotted line) and $\phi_B=90$ (green dotted line)}
\label{fig5}
\end{figure}

In Fig.~\ref{fig5}(a) and Fig.~\ref{fig5}(b), we present the relaxation rate obtained for pure Rashba and pure Dresselhaus SOI, 
respectively, at four different $\phi_B$ values.
The most remarkable finding is that $1/T_{1}$ is increased by several orders of magnitude in comparison with the case with only 
axial electric field (Fig.~\ref{fig2} and Fig.~\ref{fig3}), although strong suppressions show up at some specific combinations 
of $\phi_B$ and $\phi_E$.
 For Rashba SOI the combination is $\phi_B - \phi_E=90,270$ and for Dresselhaus SOI $\phi_B + \phi_E=0,180$.
Changes in the QD geometry do not modify significantly the qualitative results shown in Fig.~\ref{fig5}.
Only small displacements of the cancellation angles and the moderation of some minima occur.



The influence of the in-plane electric field can be explained from the fact that $E_{\parallel}$ breaks the parity 
symmetry in the direction $\phi_E$. This enables the otherwise forbidden SOI coupling between the Zeeman 
sublevels $\psi_{000}\:|\uparrow \rangle$ and $\psi_{000}\:|\downarrow \rangle$ in $\Psi_i$ and $\Psi_f$ (recall
Sec. IIIA). Since these states are very close in energy, the ensuing spin admixture is important, which
justifies the large enhancement of $1/T_1$.
In order to understand the minima we carry out a similar perturbative analysis to that of Sec. IIIA 
but now focusing on the coupling between the two $\psi_{000}$ sublevels.
Let us consider first the Dresselhaus SOI term. Assuming $H_D^{[001]} \approx H_z$ 
(as is the case for quasi-2D QDs and $\theta_z=0$), the $\phi_B=0$ matrix element is:
\begin{equation}
\langle \psi_{000} \langle \uparrow | H_z | \psi_{000} | \downarrow \rangle =
d \langle \downarrow | \sigma_y | \uparrow \rangle \langle \psi_{000} | p_z^2\,p_y | \psi_{000}\rangle
\label{pertepar0}
\end{equation}
\noindent The integral of the orbital part in Eq.~\eqref{pertepar0} vanishes when $\phi_E=0$ because of the odd parity
along $y$, but other orientations of the electric field break the parity symmetry in the $y$ direction and 
then $1/T_1$ increases, as seen in Fig.~\ref{fig5}(b) (black line).
Similar reasoning shows that for $\phi_B=90$ the parity-induced minimum occurs at $\phi_E=90$.
For intermediate magnetic field angles, however, the minimum no longer takes place when $E_\parallel \,\parallel\, \mathbf{B}$.
Indeed, for $\phi_B=45$, the minimum is found at $\phi_E=135$ ($E_\parallel \perp \mathbf{B}$). 
To explain this, it is convenient to rotate the coordinate system 45 degrees from $(x,y)$ into $(x',y')$ 
so that the $x'$ axis is aligned with the direction of $\mathbf{B}$.
As inferred from Eq.~(\ref{dressrotz}), the resulting SOI term is 
$H_z^{45}=d p_z^2 (\sigma_{y'} p_x' + \sigma_{x'} p_y')$ and the matrix element becomes:
\begin{equation}
\langle \psi_{000} \langle \uparrow | H_z^{45} | \psi_{000} | \downarrow \rangle =
d \langle \downarrow | \sigma_{y'} | \uparrow \rangle \langle \psi_{000} | p_z^2\,p_x' | \psi_{000}\rangle
\label{pertepar45}
\end{equation}
\noindent This integral vanishes due to the odd parity in $x'$ when 
$E_\parallel$ is parallel to the $y'$ axis, i.e. when $\phi_E=135$ 
in the initial coordinate frame, in agreement with Fig.~\ref{fig4}(b).

The minima in the presence of Rashba SOI can be explained in similar terms, but because $H_R^{[001]}$ 
has rotational symmetry, see Eq.~(\ref{rashrotz}), it does not change when rotating the coordinate system. 
Then, the minima always take place for $E_\parallel\,\perp\,\mathbf{B}$.

To summarize this section, the presence of in-plane electric fields greatly enhances spin relaxation
due to the lowered orbital symmetry, but the anisotropy of both Rashba and Dresselhaus SOI makes it
possible to find relative angles between $E_\parallel$ and $\mathbf{B}$ such that the relaxation
is severely reduced.

\subsection{[111] grown QDs}

\begin{figure}[h!]
\includegraphics[width=0.45\textwidth]{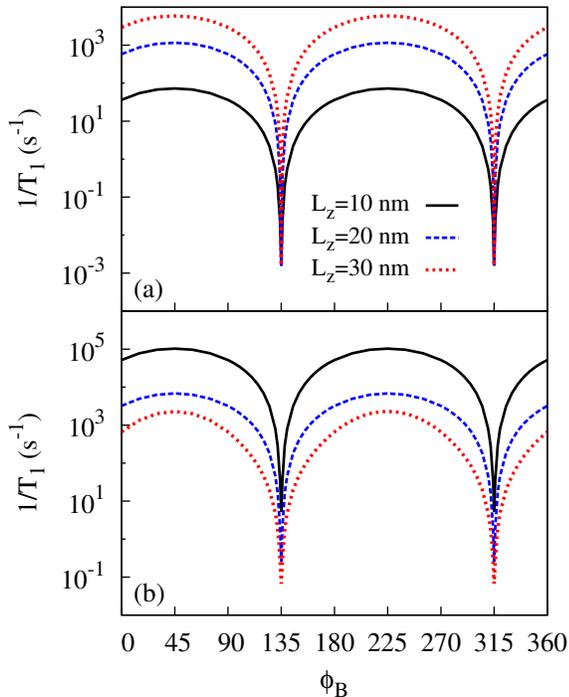}
\caption{Electron spin dynamics of square QDs grown along the [111] crystallographic direction as a function of the magnetic field orientation. Simulations considering (a) the Rashba SOI and (b) the Dresselhaus SOI are included for three QD heights: $L_z=10\:nm$ (black solid curve), $L_z=20\:nm$ (blue dashed curve) and $L_z=30\:nm$ (red dotted curve).}
\label{fig6}
\end{figure}

In Fig.~\ref{fig6} we plot the spin relaxation rate for the squared QD studied in Sec. IIIA, but now considering 
the dot is grown along the [111] crystal direction.
In general, faster relaxation rates are obtained for this orientation as compared to the [001] grown QDs.
Interestingly, we observe the same angular dependence for both Rashba SOI (Fig.~\ref{fig6}(a)) 
and Dresselhaus SOI (Fig.~\ref{fig6}(b)). Both mechanisms show strong suppressions at $\phi_B=135$ and $\phi_B=315$.
However, when increasing $L_z$ Rashba and Dresselhaus SOI mechanisms show opposite behaviors and $1/T_{1}$ 
increases and decreases, respectively. Therefore, the dot height determines which of the coupling mechanisms dominates. 

The cancellation angles of the relaxation in Fig.~\ref{fig6} can be justified noting that the canonical momenta 
$p_x=-i \hbar d/d_x +z B \sin \phi_B$ and $p_y=-i \hbar d/d_y -z B \cos \phi_B$ have exactly the same form 
for $\phi_B=135$ and $\phi_B=315$ since $L_x=L_y$.
As a result, the first term in Eq.~\eqref{rash111} and several terms in Eq.~\eqref{dress111} cancel out, 
yielding two sharp minima in the scattering rate curve.

The identical anisotropy of Rashba and Dresselhaus SOI induced spin relaxation in [111] QDs revealed by
Fig.~\ref{fig6}, which is a consequence of the formal equivalences between $H_R^[111]$ and $H_D^[111]$,\cite{Zutic2004},
 facilitates in practice the simultaneous quenching of both mechanisms. For magnetic fields
where hyperfine interaction is negligible and square dots, this should lead to spin lifetimes in the range of seconds.
We have further checked that changes in the QD base shape do not modify the qualitative behavior reported above, 
the minima being only slightly shifted for rectangular dots under Dresselhaus SOI. 

\section{Conclusions}

We have investigated systematically the electron spin scattering anisotropy in 3D cuboidal GaAs QDs 
grown along the [001] and [111] directions.  We have shown that the relaxation rate can be controlled 
by several orders of magnitude by varying the in-plane orientation of external magnetic and electric fields.

In [001] grown QDs under an axial electric field, the spin relaxation in-plane anisotropy is 
strongly dependent on the QD geometry and the nature of the dominating SOI term.
For Rashba SOI, the relaxation is isotropic or anisotropic when the base is 
squared and rectangular, respectively, and it is not affected by changes in the QD height.
On the other hand, for Dresselhaus SOI, the relaxation presents a different behavior depending not only on the base shape, 
but also on the QD height.  In fact, small and high dots can even show contrary angular dependence, 
evidencing the important role of QD three-dimensionality. 

An additional in-plane electric field component causes a strong increase in the relaxation rate, 
but certain combinations of $\phi_B$ and $\phi_E$ lead to enhanced spin lifetimes.
We find that these combinations are different for Rashba , $\phi_B - \phi_E=90,270$, 
and Dresselhaus SOI, $\phi_B + \phi_E=0,180$.

We have also shown that rotating the confinement potential in-plane with respect to the crystal 
structure causes an important modulation of the spin relaxation, that is severely suppressed 
when the high symmetry directions of the QD confinement match the main crystallographic axes. 
This modulation arises from the cubic Dresselhaus terms, which are important even for small heights.

We have further studied QDs grown along the [111] direction. We have found that Rashba and Dresselhaus 
SOI present the same angular depencence with $\phi_B$, with pronounced minima at certain magnetic
field orientations. This enables simultaneous suppression of Rashba and Dresselhaus SOI induced
spin relaxation, which is an advantadge as compared to more conventional [001] grown QDs.

\begin{acknowledgments}
This work was supported by UJI-Bancaixa Project No. P1-1B2011-01, MINECO Project No. CTQ2011-27324, and FPU Grant (C.S.).
\end{acknowledgments}

\end{document}